\documentclass[conference]{IEEEtran}

\ifCLASSINFOpdf
  \usepackage[pdftex]{graphicx}
  \graphicspath{{./graphics/}}
  \DeclareGraphicsExtensions{.pdf,.jpeg,.png,.eps,.bmp}
\else
\fi

\newcommand{\subparagraph}{}

\usepackage{multirow}
\usepackage{tabularx}
\usepackage{colortbl}
\usepackage{hhline}
\usepackage{ragged2e}
\usepackage{caption}
\usepackage{makecell}
\usepackage{amsmath}
\usepackage{amssymb}
\usepackage{txfonts}
\usepackage{color}
\usepackage{amsfonts,amssymb}
\usepackage{float}
\usepackage{titlesec}
	
\usepackage{parskip}

\usepackage[english]{babel}
\usepackage[utf8x]{inputenc}
\usepackage[T1]{fontenc}

\titlespacing{\section}{0pt}{0.5\parskip}{-0.5\parskip}
\titlespacing{\subsection}{0pt}{0.5\parskip}{-0.5\parskip}
\titlespacing{\subsubsection}{0pt}{0.5\parskip}{-0.5\parskip plus 0pt minus 0pt}
\title{Energy Expenditure Estimation Through Daily Activity Recognition Using a Smart-phone}

\author{\IEEEauthorblockN{Maxime De Bois}
\IEEEauthorblockA{University of Paris-Sud\\CNRS-LIMSI
\\Orsay, France\\
Email: maxime.debois@limsi.fr}
\and
\IEEEauthorblockN{Hamdi Amroun}
\IEEEauthorblockA{University of Paris-Sud\\CNRS-LIMSI
\\Orsay, France\\
Email: hamdi.amroun@limsi.fr}
\and
\IEEEauthorblockN{Mehdi Ammi}
\IEEEauthorblockA{University of Paris-Sud\\CNRS-LIMSI
\\Orsay, France\\
Email: mehdi.ammi@limsi.fr}}

\begin{document}

\maketitle

\begin{abstract}
This paper presents a 3-step system that estimates the real-time energy expenditure  of an individual in a non-intrusive way. First, using the user's smart-phone's sensors, we build a Decision Tree model to recognize his physical activity (\textit{running}, \textit{standing}, ...). Then, we use the detected physical activity, the time and the user's speed to infer his daily activity (\textit{watching TV}, \textit{going to the bathroom}, ...) through the use of a reinforcement learning environment, the Partially Observable Markov Decision Process framework. Once the daily activities are recognized, we translate this information into energy expenditure using the compendium of physical activities. By successfully detecting 8 physical activities at 90\%, we reached an overall accuracy of 80\% in recognizing 17 different daily activities. This result leads us to estimate the energy expenditure of the user with a mean error of 26\% of the expected estimation.

\end{abstract}



\section{Introduction}

Physical inactivity has become a major global public health problem and has been deemed responsible of 3.2 million of deaths in the world in 2004 \cite{whophys}. Being able to monitor the energy expenditure of individuals through their physical activities could help fighting modern diseases such as obesity and diabetes, which are increasingly direct consequences of physical inactivity \cite{haskell}.


The strong recent growth of the Internet of Things (IoT), with 28.1 billion of objects in 2020 \cite{iot}, opens up new ways to approach this issue.

In 2012, Altini \textit{et al.} proposed a way to estimate the energy expenditure of an individual wearing an accelerometer and a heart rate monitor on the chest \cite{altini}. First, they use regression based models to recognize clusters of physical activities. Then, using a combination of the compendium of physical activities (built by Ainsworth \textit{et al.} \cite{ainsworth}) and a pulmonary gas exchange device, they translate the physical activities into energy expenditure.

Physical activity recognition has been a dynamic field of study in the past few years. First using IoT \cite{gao} \cite{zhu}, and then using smart-phones \cite{sansegundo} \cite{anguita} \cite{amroun}. Through machine learning techniques applied to the smart-phone's accelerometer, they recognize the user's physical activities such as \textit{running}, \textit{walking} or \textit{standing}. Besides, using physical activity recognition, Weiss \textit{et al.} worked on a smart-phone application whose aim is to improve health and well-being \cite{weiss}. Recognizing those activities in real-time and for a lot of different users, they were able to give the user insights on his health (burnt calories, total time spent walking, ...).

We can highlight several limitations in those works. First, using several sensors (an accelerometer and a heart rate monitor as Altini \textit{et al} did, for instance) to estimate the user's energy expenditure is intrusive. Second, while Weiss \textit{et al.} addressed this intrusivenes issue using a smart-phone, they did not measure the energy expenditure of the user. Ainsworth \textit{et al.} showed that we need to know precisely the activity of the user (\textit{watching the TV, sitting}, instead of just \textit{sitting}) to estimate his energy expenditure \cite{ainsworth}. Thus, recognizing his physical activity (\textit{running}, \textit{standing}, ...) is not a sufficient information to compute his energy expenditure.

This paper aims at addressing those issues by presenting a non-intrusive way to estimate the energy expenditure of an individual in real time. First, using the smart-phone's embedded sensors we identify the physical activity of the user (\textit{walking}, \textit{sitting}, ...) in real-time. Then, based on this information, we recognize the daily activities of the user (\textit{going to work}, \textit{eating breakfast}, ...). Finally, with the recognized daily activity and the compendium of physical activities built by Ainsworth \textit{et al.} that assigns for every activity a metabolic equivalent, we compute the energy expenditure of the user in real-time.

\section{Methods \& Implementation}

\label{sec:architecture}

\begin{figure}
\centering
\includegraphics[scale=0.60]{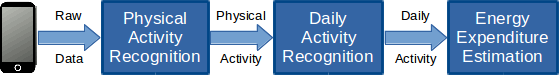}
\caption{System Flow Diagram}
\label{fig:architecture}
\end{figure}

In this section, we describe the whole methodology that makes us estimate the energy expenditure of the user in real-time. Figure \ref{fig:architecture} represents the data flow of the system and its 3-step process.

To estimate the energy expenditure of an individual, we need to know precisely what he is doing. The compendium of physical activities built by Ainsworth \textit{et al.} shows that knowing his physical activity (\textit{running}, \textit{standing}, ...) is not a sufficient information. We need to know the context of the physical activity, which we will call the \textit{daily activity}. For instance, solely knowing the physical activity, we could not differentiate those two following activities: \textit{sitting, eating} and \textit{sitting, playing the drums}. Both activities would be detected as \textit{sitting} activities, but yield very different energy expenditure values \cite{ainsworth}.


To recognize those daily activities, we chose to follow a 2-step process \cite{lee}: first we recognize the physical activity of the user (\textit{running}, \textit{standing}, ...), then we use this information to infer his daily activity (\textit{watching TV}, \textit{going to the bathroom}, ...).

Because of its computational power, its wide acceptance and since it has been shown to work well in recognizing physical activities \cite{sansegundo} \cite{weiss}, the smart-phone has been chosen as the primary input of data in this study. Besides, this system could have also worked with other IoT configurations such as with a smart-watch (alone or added to the smart-phone).




\subsection{Physical Activity Recognition}

The recognition of the physical activities of the user using his smart-phone is based on the works of Anguita \textit{et al.} \cite{anguita}, San-Segundo \textit{et al.} \cite{sansegundo} and Weiss \textit{et al.} \cite{weiss}. In those works, the following physical activities are usually recognized: lying down, climbing stairs (up or down), sitting, standing, running, walking. To do so, time-frequency features from the accelerometer (and sometimes from the gyroscope as well) are extracted. Those features are then used to build a classification model using Machine Learning algorithms (Support Vector Machine \cite{anguita}, Hidden Markov Model \cite{sansegundo}, Random Forest \cite{weiss}).\color{black}



In order to have our whole system working in real-life environments, we decided to include another activity in the recognition model: the \textit{missing} activity. This activity represents the times when the smart-phone is not with the user but left behind somewhere. Recognizing this activity makes us address the non-intrusivity of the system: we do not force the user to have his smart-phone inside his trousers' pocket.

\subsubsection{Data Collection} We chose to use the three common sensors of a smart-phone to infer the physical activity : the accelerometer, the gyroscope and the magnetometer. We developed an Android application to collect the data the following way: first, the user chooses a physical activity he wants to record (which eases the labelling of the data), then he puts his smart-phone inside his trousers' pocket and starts the recording session (lasting 2 minutes and 30 seconds) and carries out the activity. During the recording, the data of the three axes of the sensors are sampled at a 50 Hz frequency. Once done, the recorded data are saved into a local SQLite database.

\subsubsection{Pre-processing}

In order to exclude the gravitational force and to select the body movements in the signal, we filtered it between 0.3 and 20 Hz \cite{karantonis}. Then we regrouped the raw data in sliding windows of 2.56 seconds with a 50\% overlap \cite{sansegundo} \cite{anguita}. Following a methodology detailed by San-Segundo \textit{et al.}, we finally extracted the time-frequency features from the signals \cite{sansegundofeatures}. This is the list of the features we extracted :
\vspace{-2mm}
\begin{enumerate}
\item \textbf{Time domain features}: mean, standard deviation, signal magnitude area, auto-regression coefficients, energy, total energy, signals correlation.
\item \textbf{Frequency domain features}: frequency skewness, entropy, energy, total energy.
\end{enumerate}

\subsubsection{Model Building}

Once the features have been extracted, we can build the physical activity recognition model. We chose to use a Decision Tree model built on the features and the time-frequency values. We prefered this model over others (k-Nearest Neighbours, Random Forest, Support Vector Machine) because it yields the best results for little training and computing time \cite{gao}.


\subsection{Daily Activity Recognition}

As previously explained, in order to be able to estimate the energy expenditure of the user, we need to be able to understand the nature of the physical activities he is doing during the day, which we call the \textit{daily activities}. 
To do so, we built an intelligence using the \textit{Partially Observable Markov Decision Process} (POMDP) framework.


\subsubsection{The POMDP Theory}

The POMDP framework is a reinforcement learning framework where the agent, the intelligence, navigates in an environment from which it gets a partial and noisy representation. In this environment, the agent will take actions that may (or may not) affect the environment. Depending on the current state of the environment and on the action it takes, the agent receives some rewards. From trials and errors, the agent will build a predictive model.

Traditionnaly, the POMDP framework is used in robotic tasks \cite{grady}. Processing its sensors' signals, the robot builds a representation of the environment from which it can take actions.


In our work, we propose to adapt this framework to daily activities recognition. The agent tries to infer the state of the environment (the daily activities) while being given partial information of it, which we call the \textit{observation} in the POMDP framework. The pieces of observation are the following: the time, the physical activity (detected by the physical activity recognition module described earlier) and the speed of the user (computed from the GPS signals of the smart-phone). The Table \ref{table:activities} represents an example of a sequence of activities describing the user's day.


\newcommand{\specialcell}[2][c]{%
  \begin{tabular}[#1]{@{}c@{}}#2\end{tabular}}

\begin{table}[!t]
\renewcommand{\arraystretch}{1.3}
\caption{Daily Activities' Sequence Example}
\label{table:activities}
\centering
\setlength{\tabcolsep}{3pt}
\begin{tabular}{|c|c|c|c|c|}
\hline
Activity & \specialcell{Length\\(min)}  & \specialcell{Physical \\ Activity} & \specialcell{Speed \\(km/h)} & \specialcell{Corresponding\\Activity Code\textsuperscript{1}}\\
\hline
\hline
Eat breakfast & 10 & Sitting & 0 & 13030\\
\hline
Wash self & 30 & Standing & 0 & 13040\\
\hline
Get ready & 10 & Standing & 0 & 9070\\
\hline
Go to the bus & 9 & Walking & 4.5 & 17190\\
\hline
Take the bus & 8 & Sitting & 40.0 & 16016\\
\hline
Walk to work & 2 & Walking & 3.0 & 17190\\
\hline
Go upstairs & 1 & Climbing stairs & 1.0 & 17133\\
\hline
Go to the toilets & 3 & Walking & 2.5 & 17151\\
\hline
Work & 120 & Sitting & 0 & 11580\\
\hline
\end{tabular}\par
\justify
~\\
This scenario, starting at 8 a.m. and ending at 11:13 a.m., only represents a fraction of a day. Whole days are usually used to train and to test the model. ~~~~~~~~~~~~~~~~~~ \linebreak
\textsuperscript{1} See Table \ref{table:met} and see Subsection \ref{subsec:eeestimation} for more explanation.

\end{table}

More formally, a POMDP is a tuple $(S,A,T,R,\Omega,O)$ where \cite{pomdp} :
\begin{enumerate}
\item[$S$:] a finite set of states of the world, described here by all the possible daily activities of the user.
\item[$A$:] a finite set of actions, described here by all the user's daily activities' predictions.
\item[$T$:] $S \times A \rightarrow \Pi(S)$, the \textit{state-transition function}. $T(s,a,s')$ expresses the probability of getting to state $s'$ while being in state $s$ and doing action $a$.
\item[$R$:] $S \times A \rightarrow \varmathbb{R}$, the \textit{reward function}. We call $R(s,a)$ the expected immediate reward the agent is given for taking action $a$ in state $s$.
\item[$\Omega$:] a finite set of observations the agent can experience from its environment. Every observation is described here by a tuple $(time,physical\ activity,speed)$.
\item[$O$:] $S \times A \rightarrow \Pi(\Omega)$, the \textit{observation function}. $O(s',a,o)$ represents the probability of making the observation $o$ when taking action $a$ and landing in state $s'$.\\
\end{enumerate}

Since the agent is unable to observe the current state of the environment, it needs to compute what we call the \textit{belief state}, which is a probability distribution over the state space. Starting from an arbitrary initial belief state, it updates its belief of the state of the environment after taking an action $a$ and receiving an observation $o$, following \eqref{beliefupdate}.

\begin{equation}
 b'(s') = Pr(s'\mid o,a,b) = \frac{O(s',a,o) \sum\limits_{s \in S}T(s,a,s')b(s)}{ Pr(o \mid a,b)}
 \label{beliefupdate}
\end{equation}

Where $Pr(o \mid a,b)$ is a normalization factor.

The goal of the agent is to maximize the immediate expected reward expressed by the Bellman equation, which we can simplify as \eqref{immediatereward} since the agent's actions do not impact the environment (it only guesses the current state).

\begin{equation}
  V(b) = \max_{a} \left[ \sum\limits_{s \in S} R(s,a)b(s) \right]
 \label{immediatereward}
\end{equation}

It means that the agent will always predict the activity for which its associated rewards times its probability is the highest.

\subsubsection{Implementation}

All the complexity of the model revolves around the computing of the \textit{state-transition function} and the \textit{observation function}. We can derive those probabilities from experience. While this is not a problem for the transition probabilities, it is a more delicate one for the observation probabilities because of the sparsity of the data.

As a reminder, the observation function depicts the probability of having such an observation (time, physical activity, speed) given a daily activity. The issue occurs for activities that do not last long in our daily life but occur a lot (\textit{going to the bathroom} for instance). Those activities are hard to detect since they can occur at any time of the day and, most of the times, there is no observation record to support such an observation  for this daily activity.

To address the sparsity of our data, we chose to apply a technique often used in language modelling when the data is sparse \cite{deolalikar}, called \textit{Laplace smoothing} or \textit{additive smoothing}. The idea is quite simple: we add a fixed number of fake occurrences in the space we want to be smoothed and we recompute the probabilities.

However, this cannot be directly applied in our environment since the observation space is too big and the data too sparse. We then chose to tweak the Laplace smoothing technique the following way. First, for every occurrence of an observation for a given activity, we add fake occurrences to all its temporal observation neighbours. This will restrict the impact of the smoothing around the initial observation occurrences in the time axis. Second, instead of adding a fixed numbers to the neighbours of the initial occurrences, we chose to take the value on a Gaussian centred on the initial occurrence. The higher the variance of the Gaussian is, the farer from the initial observation fake occurrences are added.

\subsection{Energy Expenditure Estimation}
\label{subsec:eeestimation}

Based on the daily activity recognition module, we are now going to estimate the energy expenditure of the user.

A Metabolic Equivalent (MET) is a value that expresses the intensity of an activity. It is defined as \textit{the ratio of the work metabolic rate to a standard resting metabolic rate} \cite{ainsworth2000}. A MET equals to 1 kcal.kg\textsuperscript{-1}.h\textsuperscript{-1} (equivalent to the energy cost of sitting quietly). A MET can also be defined as the oxygen uptake in ml.kg\textsuperscript{-1}.min\textsuperscript{-1}, where 3.5 ml.kg\textsuperscript{-1}.min\textsuperscript{-1} is the oxygen uptake of an individual sitting quietly.

Over the years, Ainsworth \textit{et al.} have built a vast compendium of physical activities where for every activity they associate one code and one MET value. We will use the last update of the compendium \cite{ainsworth} to first translate the daily activities we detected in the last part (see the last column of Table \ref{table:activities}), and then to estimate, in MET, the energy expenditure over time of the user using the corresponding MET value (Table \ref{table:met}). For instance, if the user, weighing 60 kg, walks during  9 minutes to catch the bus with a walking speed of 4.5 km/h and then rides the bus during 8 minutes while sitting, we can compute the whole energy expenditure (EE) this way: $EE = 60*(9/60 * 4.5 + 8/60 * 1.3) = 51~kcal$.


\begin{table}[!t]
\renewcommand{\arraystretch}{1.3}
\caption{Sample of the Compendium of Physical Activities\textsuperscript{1}}
\label{table:met}
\centering
\setlength{\tabcolsep}{3pt}
\begin{tabular}{|c|c|c|}
\hline
\specialcell{Activity\\Code}  & Description & \specialcell{Metabolic\\Equivalent\textsuperscript{2}}\\
\hline
\hline
5035 & \specialcell{kitchen activity, general, (e.g. cooking, \\washing dishes,cleaning up), moderate effort} & 3.3\\
\hline
7025 & \specialcell{sitting, listening to music (not talking or \\reading) or watching a movie in a theater} & 1.5\\
\hline
7030 & sleeping & 1.0\\
\hline
7040 & standing quietly, standing in a line & 1.3\\
\hline
9045 & \specialcell{sitting, playing traditional video game, \\computer game} & 1.0\\
\hline
9055 & \specialcell{sitting, talking in person, on the phone, \\computer, or text messaging, light effort} & 1.5\\
\hline
9070 & standing, reading & 1.8\\
\hline
10074 & playing musical instruments, general & 2.0\\
\hline
11580 & \specialcell{sitting tasks, light effort (e.g., office work,\\ chemistry lab work, computer work, ...)} & 1.5\\
\hline
13030 & eating, sitting & 1.5\\
\hline
13040 & \specialcell{grooming, washing hands, shaving, brushing\\ teeth, putting on make-up, sitting or standing} & 2.0\\
\hline
16016 & riding in a bus or train & 1.3 \\
\hline
17070 & descending stairs & 3.5\\
\hline
17133 & stair climbing, slow pace & 4.0\\
\hline
17151 & \specialcell{walking, less than 3.2 km/h, level,\\ strolling, very slow} & 2.0\\
\hline
17152 & walking, 3.2 km/h, level, slow pace, firm surface & 2.0\\
\hline
17190 & \specialcell{walking, 4.5 to 5.1 km/h, level,\\ moderate pace, firm surface} & 3.5\\
\hline
\end{tabular}\par
\justify
~\\
\textsuperscript{1} original table is 821 entries long.\\
\textsuperscript{2} in MET or kcal.h\textsuperscript{-1}.kg\textsuperscript{-1}.
\end{table}

\section{Experimental Results \& Discussion}

This sections presents and discusses our experimental results following the 3-step methodology presented earlier: first the physical activity recognition module, then the daily activity recognition module, and finally the energy expenditure estimation.


\label{sec:results}
\subsection{Physical Activity Recognition}
The results of the physical activities recognition module have been obtained using a ten-fold cross validation, often used in activity recognition as it is a compromise between performance and computing time \cite{gao}.


Overall, across all the physical activities we achieve a performance of 90\%, ranging from 75\% (lying down) to 100\% (running). Moreover, our results, summarized in Table \ref{table:eparesults}, show that it is easier to classify dynamic activities (running, walking, climbing stairs) than static activities (sitting, laying, standing, smart-phone "missing"). While dynamic activities recognition has a performance of 98.5\%, static activities are successfully recognized with an accuracy of 82\%. Because of less intense signals' variations involved in static activities, it is harder for the model to tell one activity from another.

\begin{table}[!t]
\renewcommand{\arraystretch}{1.3}
\caption{Physical Activity Recognition Probabilities}
\label{table:eparesults}
\centering
\setlength{\tabcolsep}{2.3pt}
\begin{tabular}{c c c c c c c c c c}
{}  & {}  & \multicolumn{8}{c}{Predicted Activity}\\
\cline{3-10}
{} & \multicolumn{1}{l|}{} & lie & missing & sit & stairsdown & stairsup & stand & run & \multicolumn{1}{l|}{walk}\\
\cline{2-10}
\multirow{8}{*}{\rotatebox{90}{Expected Activity}} & \multicolumn{1}{|l|}{lie} & \textbf{0.75} & 0.06 & 0.09 & 0 & 0 & 0.1 & 0 & \multicolumn{1}{l|}{~~0}\\
\cline{2-10}
& \multicolumn{1}{|l|}{missing} & 0.1 & \textbf{0.83} & 0.07 & 0 & 0 & 0 & 0 & \multicolumn{1}{l|}{~~0}\\
\cline{2-10}
& \multicolumn{1}{|l|}{sit} & 0.09 & 0 & \textbf{0.78} & 0 & 0 & 0.13 & 0 & \multicolumn{1}{l|}{~~0}\\
\cline{2-10}
& \multicolumn{1}{|l|}{stairsdown} & 0 & 0 & 0 & \textbf{0.96} & 0.04 & 0 & 0 & \multicolumn{1}{l|}{~~0}\\
\cline{2-10}
& \multicolumn{1}{|l|}{stairsup} & 0 & 0 & 0 & 0.01 & \textbf{0.99} & 0 & 0 & \multicolumn{1}{l|}{~~0}\\
\cline{2-10}
& \multicolumn{1}{|l|}{stand} & 0.03 & 0.04 & 0.02 & 0 & 0 & \textbf{0.91} & 0 & \multicolumn{1}{l|}{~~0}\\
\cline{2-10}
& \multicolumn{1}{|l|}{run} & 0 & 0 & 0 & 0 & 0 & 0 & \textbf{1} & \multicolumn{1}{l|}{~~0}\\
\cline{2-10}
& \multicolumn{1}{|l|}{walk} & 0 & 0 & 0 & 0.01 & 0 & 0 & 0 & \multicolumn{1}{l|}{\textbf{0.99}}\\
\cline{2-10}
\end{tabular}
\end{table}

Our results are comparable to those obtained in other studies. For instance, Anguita \textit{et al.} achieved an overall performance of 89\% in the recognition of six different physical activities (walking, climbing stairs up/down, standing, sitting, lying) \cite{anguita}. As for Weiss \textit{et al.}, they obtained a global performance of 95\% in detecting five different activities: jogging, stairs, walking, standing and sitting/lyings down. The difference in the prediction success rate is explained by the grouping of the \textit{lying} and the \textit{sitting} activities together (9\% of confusion between those two activities, see Table \ref{table:eparesults}) and that they do not recognize the \textit{missing} activity.

\subsection{Daily Activity Recognition}
To understand the nature of the physical activities we just recognized, we trained an intelligence in a POMDP environment using the user's daily sequences of activities. The results, depicted in Table \ref{natureresults} are obtained using a ten-fold cross-validation: nine days of activities are used for training, and the remaining day is used for testing. The performance of the system has been evaluated in two different ways: the mean performance of the activities' recognition, and the weighed mean performance of the activities' recognition, with the weight depending on the length of the activities (the longer, the higher the weight). Three kinds of experiment have been tested: the base POMDP model, the base POMDP model where we smoothed the observation probability function, and the base POMDP model with smoothed observation probability function and tweaked rewards.

\begin{table}[!t]
\renewcommand{\arraystretch}{1.3}
\caption{Daily Activity Recognition Results}
\label{natureresults}
\centering
\setlength{\tabcolsep}{3pt}
\begin{tabular}{|c|c|c|c|c|}
\hline
Experiment Type & Mean & Weighed Mean & Min & Max\\
\hline
\hline
Base & 0.45 & 0.76 & 0 & 0.98\\
\hline
With smoothing & 0.79 & 0.75 & 0.002 & 0.98\\
\hline
\specialcell{With smoothing \& \\tweaked rewards} & 0.80 & 0.78 & 0.29 & 0.99\\
\hline
\end{tabular}\par
\justify
\end{table}

While the performance over the day (weighed mean) is quite the same for the three experiments, we can notice a big variation in the mean performance and in the minimum prediction success rate.

First, without the smoothing of the probabilities presented earlier, the mean performance is 45\%. Long activities (such as \textit{sleeping}) are easily detected increasing the weighed performance. In the other hand, a lot of short activities (such as \textit{going to the toilets}) have a recognition success rate of 0\%. Smoothing the observation function increases this performance above 79\% : 80\% of the activities have a recognition rate of at least 80\%.

Because some activities have still a low recognition accuracy (\textit{playing video games on the computer} with 0.2\%, \textit{watching a movie on the computer} with 37\%, ...), we tried to increase the rewards associated with those activities' predictions, which is represented by the \textit{with smoothing \& tweaked rewards} exeriment type in Table \ref{table:eparesults}. With a higher reward, we encourage the agent to choose this activity as the prediction even if the odds are not very good. This improved our results, by slightly increasing the mean and weighed mean performance, but most of all, by increasing the minimum recognition success from 0.2\% to 29\%.

\begin{figure*}
\centering
\includegraphics[scale=0.69]{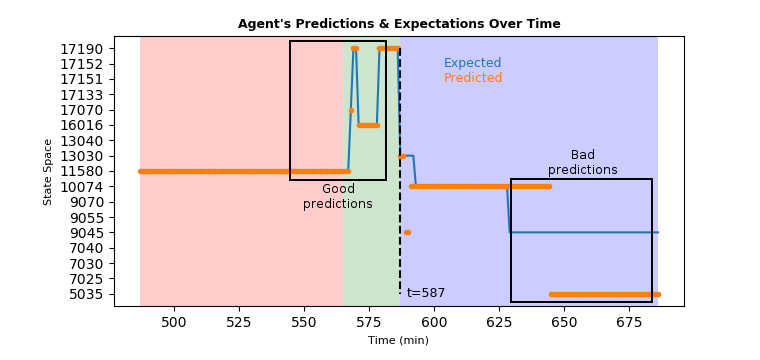}
\caption{Daily Activity Recognition, Predicted VS Expected over Time\color{black}}
\label{activitiesnatureovertime}
\end{figure*}

Since the protocols in daily activity recognition vary heavily between studies, we cannot compare our results to others. However, there are still things to be noted about our results. The activities that have the lowest performance happen during the evening. During this time, the user is at home and does not have his smart-phone in his pocket. The physical activity detected by our physical activity recognition module is then the \textit{missing} activity. When the physical activity detected by the smart-phone is not \textit{missing}, the success in the prediction is about 98\% against 61\% when it is. The Figure \ref{activitiesnatureovertime} represents this shift of performance. The red area corresponds to working activities around 4:00 pm. Then the user goes back home, which corresponds to the activities in the green area. Until he is home, which is represented by the point \textit{t=587},  the daily activity recognition is close to 100\% success. Afterwards, the user performs home activities such as \textit{eating}, \textit{playing the guitar}, \textit{playing on the computer}, leaving his smart-phone behind. This shows the importance for the intelligence to know the physical activity of the user. Without it, it cannot tell one daily activity from another.

\subsection{Energy Expenditure Estimation}

The final step of this system is the translation of the daily activities recognized in real-time (every minute) into energy expenditure. To do so, every activity has been associated an activity code in the compendium of physical activities built by Ainsworth \textit{et al.} \cite{ainsworth}. For every activity code corresponds one metabolic equivalent value (in MET), see Table \ref{table:met}. The results, displayed in Table \ref{eeresults} have been obtained choosing the \textit{with smoothing \& tweaked rewards} experiment type for the recognition of daily activities.

\begin{table}[!t]
\renewcommand{\arraystretch}{1.3}
\caption{Energy Expenditure Estimation Results}
\label{eeresults}
\centering
\setlength{\tabcolsep}{3pt}
\begin{tabular}{|c|c|c|c|}
\hline
\specialcell{Mean Absolute\textsuperscript{1}} & \specialcell{Mean At the\\ End of the Day\textsuperscript{1}} & Min\textsuperscript{1} & Max\textsuperscript{1} \\
\hline
\hline
26\% & +20\% & -70\%\textsuperscript{2} & +230\%\textsuperscript{2}\\
\hline
\end{tabular}\par
\justify
~\\
\textsuperscript{1} The values are expressed in percentage of the difference with the expected values.\\
\textsuperscript{2} Both values are obtained when the intelligence confuses the \textit{washing dishes} (moderate effort) activity with the \textit{playing video games} (light effort), and vice versa.
\end{table}

\begin{figure}[H]
\centering
\includegraphics[scale=0.60]{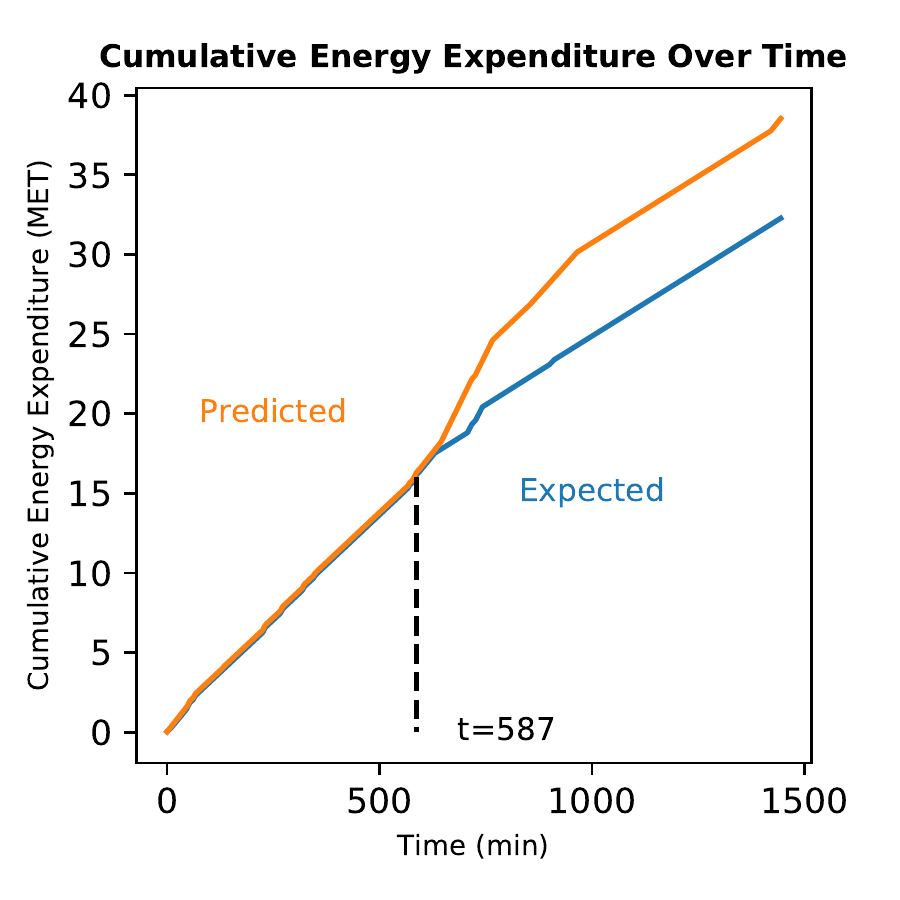}
\caption{Cumulative Energy Expenditure Over Time}
\label{energyexpenditureovertime}
\end{figure}

The Figure \ref{energyexpenditureovertime} represents the expected (in blue) and the predicted (in orange) cumulative energy expenditure over time. As for the daily activity recognition module, until the time of the day \textit{t=587}, where the user comes back home and leaves his smart-phone behind, the energy expenditure prediction is close to the expected value. However, the energy expenditure is wrong afterwards since the daily activity recognition is not as accurate.

\section{Conclusion \& Future Works}
\label{sec:conclusion}

In this study, we built a non-intrusive system using a smart-phone that estimates the energy expenditure of its user in real-time. The system uses a two-step activity recognition module. First it detects the physical activity of the user (\textit{running}, \textit{sitting}, ...) from the smart-phone sensors. Then, with this information, the system infers the daily activities of the user (\textit{watching TV}, \textit{going to the bathroom}, ...), which is then used to compute his real-time energy expenditure.


As also shown in previous works in this field, with 90\% accuracy,  physical activity recognition works well using the smart-phone's sensors. It is not as easy to understand the nature of the detected physical activities, though. We showed that with appropriate methods (probabilities smoothing) to overcome the sparsity of the data we are able to recognize the daily activities in real time with an overall performance of 80\% over the activities. By successfully recognizing the user's activities in real-time, we propose a way to estimate his energy expenditure in real-time using the compendium of physical activities built by Ainsworth \textit{et al.}. We achieve a mean error of 26\% of the expected energy expenditure estimation.

However this successfulness depends heavily on the activities' recognition success. We showed that, due to the user's habits, the smart-phone may not be able to detect his physical activity. When it happens, the performance drops from 98\% to 61\% which directly impacts the energy expenditure estimation. To overcome this issue, we plan on using a smart-watch instead of a smart-phone in future studies. Weiss \textit{et al.} have shown that using a smart-watch in physical activity recognition is more efficient than a smart-phone \cite{weisswatch}, especially in detecting hands activities (drinking, eating, ...). Besides, if the user has a day that is very different from the days the model has been trained on (e.g. the user has taken a day off while the model has been trained on working days), the model will have a very hard time correctly estimating the energy expenditure. If such situations are detected, we could work on modifying the estimation of the energy expenditure accordingly.

Finally, since we use a generic model to estimate the energy expenditure (the compendium of physical activities), in the future, we could try to evaluate the relevance of the model using a pulmonary gas exchange device during the experiments.

\section*{Acknowledgment}
This work is supported by the "IDI 2017" project funded by the IDEX Paris-Saclay, ANR-11-IDEX-0003-02
\setlength{\parskip}{0.25\baselineskip plus2pt minus2pt}

\end{document}